\documentclass[twocolumn,preprintnumbers,amsmath,amssymb]{revtex4}

\usepackage{psfrag}
\usepackage{graphicx}
\usepackage{dcolumn}
\usepackage{bm}
\usepackage{epsfig}
\usepackage[notcite,notref]{}

\begin{document}

\title{Self-Induced Decoherence in the Bose-Hubbard model}

\author{L. Rico-P\'erez}
\email{rico@rhrk.uni-kl.de}
\affiliation{Physics Department, University of Kaiserslautern,
Erwin-Schr\"odinger-Stra\ss e. 46, D-67663 Kaiserslautern, Germany}


\begin{abstract}
The conventional conception of decoherence relies on the interaction with an external set of degrees of freedom - the `bath' - to which the system loses quantum information. But the role of the bath can be played too by any internal degrees of freedom that are not accessible to the observer, and in this sense we can talk about `Self-Induced Decoherence'. Simulating the exact time-evolution of the few-body Bose-Hubbard Hamiltonian, we provide numerical evidence of the decay of quantum purity in the sense of `temporal typicality' referred in the context of pure state Quantum Statistical Mechanics. We analyse the causes of such purity loss in terms of the structural differences of the interacting many-body Hamiltonian in comparison with its non-interacting counterpart, finding that the predominant role in the long term behaviour is played by the shifts in the many-body energy spectrum.
\end{abstract}

\maketitle


\section{Introduction}
\label{section:Intro}

The study and explanation of quantum decoherence is necessary from both a philosophical and a practical point of view. On one hand, because the world we see in our everyday life follows classical rules rather than quantum ones and the emergence of one from the other is a fundamental issue concerning our conception of physical reality. And on the other hand, because plenty of modern research, like the recent advances on quantum computation, are completely based in the quantum framework and holding  a map that describes in detail the frontiers of the quantum realm is not optional.\\
Most extended treatments of quantum decoherence are based on the presence of an external agent, the `bath', that absorbs information from the system \cite{Zurek2003}. This is, quantum entanglement with external degrees of freedom limits the quantum behaviour of the system under study, forcing it to behave in a way that resembles the classical rules. This conception is correct but, if na\"ively understood, it may look as if classical behaviour of a system would be impossible to reach without the external assistance. Such interpretation, rather than answering the questions about the origin of decoherence, passes it further to the bath without solving it \cite{Kastner2014}.\\

This paradox is easily solved if we abstract the definition of the bath. This role can be played by any degrees of freedom, either external to the system or internal, that are not accessible to the observer. In this sense we can distinguish between the more traditional approach - the Environmentally-Induced Decoherence (EID) - caused by external agents, and the less apparent one, caused even when the system is isolated by degrees of freedom within it that cannot be traced - the Self-Induced Decoherence (SID) \cite{Castagnino2005}. The second definition can be very useful in the context of many-body theory because if our model of the system is, as usual, limited to its one- or two-body description while the particle number is much larger, then even when the system is isolated we can expect decoherence due to the coupling of our coarse-grained picture of the system to correlations of high order that we cannot measure.\\
Therefore, the statement claiming that an isolated quantum system cannot present decoherence would not be fully correct. Reflecting on this affirmation we can stablish an analogy to another very similar one extracted from the context of Classical Statistical Mechanics: the contradiction between the reversibility of microscopic dynamics (and its quasi-cyclical evolution, according to Poincar\'es Theorem) against the inevitable irreversibility established by the Second Law of Thermodynamics. Obviously, this contradiction is only superficial but reconciling microscopical reversibility/cyclicality with thermodynamic irreversibility requires an effort reconsidering the differences between our macroscopic picture of the system, ruled by the Second Law, and the microscopic model, that is time-reversible. Given the resemblance between both questions we will use the same conceptual tools used in the context of pure-state Quantum Statistical Mechanics \cite{Gogolin2016}, closely related to ergodicity and typicality \cite{Goldstein2010}.\\

This  work is structured as follows. First, in Section \ref{section:SID}, we will present the concept of Self-Induced Decoherence, first explaining that it refers to the coarse-grained description of a many body system in \ref{subsection:CGRepresentations}, then we will present quantum purity as our quantitative guideline to determine if a system has undergone a decoherence process in \ref{subsection:PurityAsQuantifier} and we will add a short remark on temporal typicality as a core idea in our theory, inherited from Statistical Mechanics, in \ref{subsection:TimeAveraging}. Next, in Section \ref{section:Analysis}, we will not start by analysing the causes of decoherence - but the causes of coherence. In other words, we will shortly  reflect on why non-interacting systems keep quantum purity constant. From there, in \ref{subsection:EDandES}, we will compare the roles of the two characteristics of interacting systems that deprives them from that feature of ideal systems: the Eigenstate Deformations and the Energy Shifts. Finally, we will further explore the role of the Energy Shifts in \ref{subsection:PurityEvolution} to estimate how much does the amplitude of purity fluctuations decrease as we increase the size of the system. We will end summarizing our conclusions in \ref{Conclusions}.\\


\section{Self-Induced Decoherence in the many-body context }\label{section:SID}

Literature related to quantum decoherence \cite{Zurek2003} departs almost invariably from the following  assumption  - that the system considered in some sense in contact with external agents. This can be either because of some interaction similar to friction \cite{Caldeira1983}, or because of the absorption and emission of electromagnetic radiation \cite{Pfau1994} or even because we are literally studying an open system that can exchange particles with another subsystem with known properties. Whatever the nature of such external agents is they are the ultimate responsible of the loss of quantum coherence that the system may undergo. While this point of view is legitimate and explains the vast majority of real situations, where the total isolation of the system is not possible, we may keep the feeling, also legitimate, that the question about what is the origin of decoherence has not been answered, but displaced instead: we have not explained how quantum systems stop behaving coherently, we have only discharged any responsibility upon `external agents', eluding to provide an actual answer \cite{Kastner2014}.\\

This dilemma has of course a solution. Nevertheless, we cannot properly understand it without turning back to the same foundational principles of Statistical Mechanics that allows to explain thermal equilibration in isolated systems. We now make a short review of these concepts.\\

\subsection{Coarse-grained description of quantum systems}\label{subsection:CGRepresentations}

We depart from this assumption, that a quantum system of $N$ particles can be described in terms of an N-body wave function  $|\Psi_N\rangle$ and the N-body density matrix will be

\begin{equation}\label{eq:NPureState}
\hat \rho_N = |\Psi_N\rangle \langle \Psi_N|
\end{equation}

While the treatment could be generalized to non-pure N-body density matrices we will only address here systems that are pure in its full quantum representation, i.e. at the N-body level of description. One of the reasons why this is not necessary is because any model and measurements concerning a quantum many-body systems will be restricted anyway due to the limited amount of information about it that we can actually access and manage. Describing a N-particle wave function is basically impossible as soon as we hit a not so high number of particles due to the fast growth of the dimensionality of the Hilbert space required to represent it when N becomes large.\\

The conventional strategy is to work with reduced representations of the system, i.e. coarse-graining our model of it. Such representations are only as complicated as the one- or two-body correlations at most. For example, we can formally define the following correlation functions

\begin{eqnarray}
G_1(\vec r| \vec r^\prime) &= & \langle \hat \psi^\dagger(\vec r)\, \hat \psi(\vec r^\prime) \rangle\\
. \, . \, .& & \nonumber\label{eq:Traditional1BodyCorrelations} \\
G_n(\vec r_1, ..., \vec r_n| \vec r_1^\prime, ..., \vec r_n^\prime) & =&  \left\langle \prod_{j=1}^n\hat \psi^\dagger(\vec r_j)\,\prod_{k=1}^n \hat \psi(\vec r_k^\prime) \right\rangle\label{eq:TraditionalNBodyCorrelations}
\end{eqnarray}

If we intend to work in terms of values {\it per particle} it is convenient to define an artifact that provides the same information than these correlation functions but `scaled'. We can then use the {\it n-body Reduced Density Matrices}, that can be defined recursively in the context of quantum systems of identical particles as follows

\begin{equation}\label{eq:DefinitionPartialTrace}
\hat \rho_{n-1} = \mbox{Ptr}_n (\hat \rho_n) = \frac{1}{n}\sum_{j=1}^M \hat a_j\, \hat \rho_n\, \hat a_j^\dagger
\end{equation}

These Reduced Density Matrices correspond to a (fictitious) system of $n$ particles that presents the same expectation values `per particle' as the actual N-body system  that we are really studying, so that

\begin{equation}\label{eq:QuantumIntensiveVariables}
G_n(\vec r_1, ..., \vec r_n| \vec r_1^\prime, ..., \vec r_n^\prime) = {{N}\choose{n} }\langle \vec r_1, ..., \vec r_n |\hat \rho_n| \vec r_1^\prime, ..., \vec r_n^\prime \rangle
\end{equation}

As long as $n\ll N$ using the reduced representation is considerably simpler. In the present work we focus on $n=1$. In general, since the access to information corresponding to correlations of very many particles is strongly limited, it is not common to consider $G_n$ for $n>1$.

\subsection{Purity as a quantifier of quantum coherence}\label{subsection:PurityAsQuantifier}

The next aspect we must consider is when can we claim that the system has lost quantum coherence. Under the previously exposed assertion, i.e. that all the information of the system available is contained within its coarse-grained representation, the most natural criterion is to accept that a quantum many-body system will be coherent if its coarse-grained representation is. But once we accept this a less obvious aspect requires to be determined as well, and this is how are we going to quantify the degree of coherence of the system.\\

In an experimental context, for example, quantum decoherence manifests though the loss of imaging contrast in the interference patterns between two or more drops of a coherent gas sample, like a Bose-Einstein Condensate \cite{Ketterle1997,Ott2004}. But since our analysis is theoretical we need a more fundamental quantity, more general and simpler to calculate. Given the relation between decoherence and entanglement it makes sense to consider some form of entropy, which is used in the context of quantum information theory as an acceptable quantifier of the degree of entanglement between the system and either external agents or unobserved degrees of freedom. While there are several available forms and definitons of entropy we will make use here of the simplest possible from the algebraic point of view: the Renyi entropy of index two, $S_{2}$, also called linear entropy 
\\

\begin{eqnarray}\label{eq:RenyiEntropy}
S_{\nu} &=& \frac{1}{1-\nu} \, \log{\mbox{Tr}\left( \hat \rho^\nu \right)}
\end{eqnarray}

where we have made use of the natural representation of the on-body reduced density matrix\\

\begin{equation}\label{eq:NaturalRepresentation}
\hat \rho_1 = \sum_{\alpha}p_\alpha \, |\alpha\rangle \langle \alpha|
\end{equation}

The greatest advantage of this particular choice is that it is specially simple to calculate because it is directly related to the so called quantum purity $P$ \\

\begin{eqnarray}\label{eq:DefPurity}
P &=& e^{-S_2} = \mbox{Tr}\left( \hat \rho_1^2\right)= \sum_{p,q=1}^M |\rho_1(p|q)|^2
\end{eqnarray}

The latter one requires no explicit diagonalization of the reduced density matrix to calculate it. We will use it in this work as a rough quantifier of the degree of decoherence achieved by our test system after undergoing a relaxation process.

\subsection{Time-averaging and its role in relaxation}\label{subsection:TimeAveraging}

Our concept of coherence loss will be based in this idea, that quantum purity will show the tendency to evolve towards low values. But against this criterion the following objection could be argued - that it is not guaranteed that for a given system any quantity that we measure (except for a few constants of motion)  will evolve towards a constant or merely stable value.\\
Even more, in the case of quantum systems with a discrete energy spectrum any amount that we observe is always a sum of periodic terms, which frequencies are given by the Hamiltonian eigenvalues. Therefore, any measured quantity will return to its initial value or a very similar one after a long enough time lapse. Although it is not formally a quantum observable, quantum purity would be no exception.\\

This question is not new. A very similar one was posed in the context of Classical Statistical Mechanics, where Poincar\'e's  Theorem ensures the return to conditions similar to the initial ones after a long enough time lapse. In that case one of the basic aspects to formulate the theory in a consistent way was not only to coarse-grained the description of the system to a macroscopic representation of it - but also to abandon the instantaneous values of the variables considered $A(t)$ in favor of its time-averaged counterparts $\bar A$

\begin{eqnarray}\label{eq:DefTimeAverage}
\bar A(T) = \frac{1}{T}\int_0^t dt^\prime\, A(t^\prime)
\end{eqnarray}

and that for practical purposes we could consider that the system evolved fast enough to take the infinite time limit $\bar A = \lim_{T\rightarrow \infty} \bar A(T)$.\\
The argument is that the time scales considered when we analyse the system from a macroscopic point of view are significantly larger than the time scales relevant to describe its microscopic evolution. Then, even admitting that the system is under constant evolution, it would not make much sense to examinate carefully the instantaneous values and it would be more useful to take the averaged value $\bar A$.\\

Once we admit that $\bar A$ should be considered the physically relevant quantity, one of the foundations of most of the formalism of Statistical Mechanics is to assume that the system considered  will hold the Ergodic Hypothesis. This is, that given enough time, we can consider that it will evolve through all of the available region of phase space uniformly. Despite being a core aspect of the standard formalism, the Ergodic Hypothesis is far from being a settled question  \footnote{Let this textbook quotation be a sample: {\it "The Ergodic Theorem has so far been an interesting mathematical exercise irrelevant to physics"}. From K. Huang, {\it Statistical Mechanics} (2nd Edition), John Wiley \& Sons Inc. (1987)}. In the quantum context such controversy only gets worse: for example, the first version of the Quantum Ergodic Theorem from J. von Neumann \cite{vonNeumann} was for a long time misinterpreted \cite{Goldstein2010} and the thermalization of isolated quantum systems has been the subject of very recent discussions \cite{Rigol2008, Cramer2008,Rigol2012,Gogolin2016}. Although these are fascinating questions, we refrain from arguing here whether the values after relaxation do coincide with those supplied by the Ergodic Hypothesis or not, or if they correspond to the values of a state of thermal equilibrium. We will be satisfied as long as the value of quantum purity after a relaxation process can be considered stable.\\

This is precisely the aspect that we should most worry about: even accepting that the quantities evaluated, purity $\bar P$ in our case, do `average away', we keep the uncertainty of having spontaneous revivals. Does this mean that a system that departs from a highly coherent state and undergoes a relaxation process that makes it lose its coherence could return to its initially coherent state given enough time? If such revivals do happen, does it still make sense to talk about decoherence even then? The answers are, respectively, it depends and yes.\\
Just as for a classical system, the question is not simply about fulfilling the Ergodic Hypothesis at the limit $T \rightarrow \infty$. We should wonder instead if the time required for the system to fulfil the hypothesis to a reasonable degree of precision is not too long; and how frequently and in what magnitude do the instantaneous values deviate from the time averaged one. In the latter sense, it is more meaningful to pay attention to the amplitude of the deviations

\begin{eqnarray}\label{eq:DefAverageFluctuations}
\bar \sigma_{A}^2(T) = \frac{1}{T}\int_0^t dt^\prime\, \left(A(t^\prime) - \bar A(T) \right)^2
\end{eqnarray}

The most convincing result in Statistical Machanics is precisely that for many systems the amplitude of such deviations is of order $\sigma_x \approx O(V^{-1/2})$ for intensive variable (i.e. for values {\it per particle}) $x=\frac{X}{V}$ when the system is very large.\\

In our case the parameter of interest is quantum purity. Being a non-linear function of the reduced density matrix, purity is not strictly a quantum observable and we must expect different behaviour in comparison with actual observables but we can be guided by exactly the same general criterion. When a quantum many-body system departs from an initial state with a coherent coarse-grained representation, we expect the corresponding quantum purity to be high and we will say that the system loses coherence if the time-averaged value $\bar P$ decays after some time to a value comparatively smaller. And we expect the value of purity fluctuations $\bar \sigma_P$ to be very small in comparison with the amount of purity loss. We can guess too that the amplitude of the fluctuations will drastically decay as we consider larger and larger systems.\\



Given the arguments exposed so far, it should be out of discussion already that Self-Induced Decoherence should be seriously considered as an alternative mechanism to Environmentally-Induced Decoherence. Nevertheless, we should still discuss in further depth the characteristics of a system that allow the phenomen, and consequently, which of its attributes require more attention if we are to predict at best this behaviour.

\subsection{The Bose-Hubbard Hamiltonian as a test system}\label{subsection:IntroBH}

In the following sections we will use the Bose-Hubbard Hamiltonian to illustrate the principles that we are explaining. We made this particular choice because it is simple to find numerical exact solutions for a few particles and because it has been studied exhaustively analytically, numerically and experimentally (as a model for one-dimensional lattices in magneto-optical traps) \cite{Bloch2005}.\\

\begin{eqnarray}\label{eq:DefBoseHubbardHamiltonian}
\hat H_{BH} &=& \sum_{m=1}^M \left( \hat a^\dagger_{m} \hat a_{m+1} + \hat a^\dagger_{m+1} \hat a_{m}+ V \, \hat n_m (\hat n_m-1)\right)\nonumber\\
& & 
\end{eqnarray}

If this form is unfamiliar to the reader it is only because we used $\hbar = J = 1$ and $V=U/J$.\\

For our simulations we will take periodic boundary conditions and a number of lattice sites $M$ and a number of particles $N$ quite limited, $N,M\leq 6$. The figures will focus on the case $N=M=5$ and $V=1/4$ to avoid unnecessary redundancy. Although the systems so modelled are quite small we will find out that the principles here exposed are already fulfilled in these cases.\\

We will also choose a concrete initial state for our numerical simulations, one that has all of the particles in one lattice site at position $x$

\begin{eqnarray}\label{eq:InitialStateSingleModeBH}
|\Psi_N^0\rangle &=& \frac{(\hat a^\dagger_{x=1})^N}{\sqrt{N!}}|0\rangle = |n_1=N,n_{j\neq 1}=0\rangle
\end{eqnarray}

By exact diagonalization of this Hamiltonian we can show the N-body evolution of this particularly simple initial state at practically any time.


\section{Analytical description of Self-Induced Decoherence}
\label{section:Analysis}

To understand the causes of Self-Induced Decoherence from an analytical point of view our strategy is to formulate the opposite question. Given the vast amount of possible N-body Hamiltonians that could model a physical system, why would any Hamiltonian in particular conserve the quantum purity of the coarse-grained representation of the system? If we knew nothing about how the many-body Hamiltonian has been formulated, if we did not know that it has been written using the second quantization formalism out of one- or two-particle terms, and required only the Hamiltonian to be a hermitian operator in the N-particle space, the we would consider the conservation of coarse-grained quantities to be a very formidable and counter-intuitive feature that only a few Hamiltonians may fulfil. The key is first of all to understand that a Hamiltonian that does so should be the exception, not the rule. And once we determine the characteristics that allow such Hamiltonian to behave in this anomalous way, we may remove them one by one to evaluate the role of each.\\

The obvious shortcut offered by this strategy resides in the fact that we do indeed know a set of Hamiltonians that do conserve the quantum purity of the reduced representation of the system. They are of course the non-interacting Hamiltonians. In its second quantization representation they have the form

\begin{eqnarray}\label{eq:IdealHamiltonian2ndQuantization}
\hat H_0 &=& \sum_{k=1}^M \hbar \, \omega^0_k \, \hat a_k^\dagger \hat a_k 
\end{eqnarray}

where $k=(1,...,M)$ are the $M$ energy eigenstates of the system when there is only one particle. Since we are not interested in open systems we will not work in second quantization any further and we will instead restrict our analysis to the Fock layer of $N$ particles, so that our Hamiltonian (restricted to that Fock layer) can be written as

\begin{eqnarray}\label{eq:IdealHamiltonianNBody}
\hat H_0^{(N)} &=& \sum_{\vec n \,/\, N,M} \hbar\, \omega_{\vec n}^0| \omega_{\vec n}^0\rangle  \langle \omega_{\vec n}^0 |\\
&=& \sum_{\vec n \,/\, N,M} \hbar\, \vec n \cdot \vec \omega^0\,| \vec n\rangle  \langle \vec n |
\end{eqnarray}

where the ideal eigenvalues are expressed as

\begin{eqnarray}\label{eq:IdealHamiltonianEigenvalues}
\omega_{\vec n}^0 &=& \vec n \cdot \vec \omega^0 = \sum_{p=1}^M n_p \omega_p^0
\end{eqnarray}

and the ideal eigenstates are

\begin{eqnarray}\label{eq:IdealHamiltonianEigenstates}
|\omega_{\vec k}^0\rangle &=& | \vec k \rangle
\end{eqnarray}

We have used the following non-conventional notation for the sums to shorten the expressions

$$ \sum_{\vec n\, /\, N,M} \rightarrow \sum_{n_1+n_2+...+n_M = N}$$

this is, $\vec n / N, M$ denotes all distributions of $N$ particles among $M$ single-particle levels, i.e. all vectors $\vec n = (n_1,n_2,...,n_M)$ with $\sum_{m=1}^M n_m = N$. We express in a similar way the vector of non-interacting eigenvalues $\vec \omega^0 = (\omega^0_1,...,\omega^0_M)$.\\

If we know the Hamiltonian in its diagonal form it will be simple also to determine the state of the system in any future instant by using the N-body time evolution operator

\begin{eqnarray}\label{eq:IdealTimeEvolutionOperator}
\hat U_0(t) &=& \sum_{\vec n \,/\, N,M} e^{-i\,  \omega^0_{\vec n}}\,| \omega_{\vec n}^0\rangle  \langle \omega_{\vec n}^0|\\
& =& \sum_{\vec n \,/\, N,M} e^{-i\, \vec n \cdot \vec \omega^0}\,| \vec n\rangle  \langle \vec n |
\end{eqnarray}

Taking any initial N-body state and the above described operator we can iterate the partial trace operation (\ref{eq:DefinitionPartialTrace}) or use the conventional one-body correlations (\ref{eq:Traditional1BodyCorrelations}) until we obtain the one-body representation at that instant and we can see that its matrix elements in the one-body energy basis are

\begin{eqnarray}
\rho_1(p|q) &=& \frac{1}{N}\langle \Psi_N(t)|\hat a_p^\dagger \hat a_q | \Psi_N(t) \rangle\nonumber\\
&=& \sum_{\vec n \, /\, N-1,M} \frac{\sqrt{(n_p+1)(n_q+1)}}{N}\nonumber\\
& &\times \langle \vec n q| \,e^{-i \frac{t}{\hbar}\, \hat H_0 } |\Psi_N^0 \rangle \langle \Psi_N^0 | e^{i \frac{t}{\hbar}\, \hat H_0 } \,| \vec n p\rangle\label{eq:IdealEvolution1RDMnone}\\
&=& \sum_{\vec n \, /\, N-1} e^{-i t (\omega^0_{\vec n p} - \omega^0_{\vec n q})}\, \frac{\sqrt{(n_p+1)(n_q+1)}}{N}\nonumber\\
& &  \langle \vec n p | \Psi_N^0\rangle \langle \Psi_N^0 | \vec n q \rangle\label{eq:IdealEvolution1RDMstates}\\
&=& e^{-i t (\omega^0_{p} - \omega^0_{q})}\, \sum_{\vec n \, /\, N-1}  \frac{\sqrt{(n_p+1)(n_q+1)}}{N}\nonumber\\
& &  \langle \vec n p | \Psi_N^0\rangle \langle \Psi_N^0 | \vec n q \rangle\label{eq:IdealEvolution1RDMenergies}\\
&=& e^{-i t (\omega_p^0 - \omega_q^0)}\,\rho^0(p|q)\label{eq:IdealEvolution1RDM}
\end{eqnarray}

where we used the simplified notation $\sqrt{n_p+1}|\vec n p \rangle = \hat a_p^\dagger | \vec n \rangle $.\\

Logically, we have obtained the same result that one could expect from the ideal Hamiltonian - that every matrix element in the one-body energy representation will only change by a phase factor $e^{-i t (\omega_p^0 - \omega_q^0)}$. The interesting step now is to pay attention to how this simplification was achieved. It has been so because of two happy coincidences that may not have to take place in an arbitrary N-body hermitian operator. One of them happens as we pass from (\ref{eq:IdealEvolution1RDMnone}) to (\ref{eq:IdealEvolution1RDMstates}), and it happens because $\langle \vec n p| \omega^0_{\vec m}\rangle = \delta_{\vec m, \vec n p}$. The second one happens as we pass from (\ref{eq:IdealEvolution1RDMstates}) to (\ref{eq:IdealEvolution1RDMenergies}), because $\omega^0_{\vec n p} - \omega^0_{\vec n q} = \omega^0_p - \omega^0_q$. These are the two features of the non-interacting Hamiltonian that allow it to conserve the purity of its coarse-grained representation. Now we will proceed to evaluate their individual roles.

\subsection{Roles of Energy Shifts and Eigenstate Deformation}\label{subsection:EDandES}

Among all possible N-body Hamiltonians that one may take the non-interacting ones are the exception, not the rule. Consequently, we must now study the effect of adding interactions. To make this analysis easier we will assume that this interaction is weak enough to consider that we are not `far' from the ideal case. This means that, departing from a known non-interacting Hamiltonian $\hat H_0^{(N)}$ as described in (\ref{eq:IdealHamiltonianNBody}), its interacting counterpart $\hat H^{(N)}$ can be written in the following form

\begin{eqnarray}\label{eq:defEDandES}
\hat H^{(N)} &=& \sum_{\vec k\, /\, N} \hbar \omega_{\vec k} \, |\omega_{\vec k}\rangle \langle \omega_{\vec k} |\\
\omega_{\vec k} &=& \vec k \cdot \vec \omega^0 + \Delta_{\vec k} \\
|\omega_{\vec k} \rangle &=& |\vec k \rangle + |\phi_{\vec k}\rangle
\end{eqnarray}

This is, as we apply the interaction, we cause each N-body energy level $\omega_{\vec n}$ to be displaced by a quantity that we will call {\it Energy Shift} $\Delta_{\vec n}$, and that the natural N-body basis of the Hamiltonian suffers a unitary transformation that displaces each eigenvector according to $|\omega_{\vec n}\rangle = |\omega_{\vec n}^0\rangle + |\phi_{\vec n}\rangle$. We will refer to $|\phi_{\vec n}\rangle$ or its components $\phi_{\vec m,\vec n}=\langle \vec m | \phi_{\vec n}\rangle$ as the {\it Eigenstate Deformation}. As we have seen above when we calculated (\ref{eq:IdealEvolution1RDM}) each one of these two attributes correspond to a qualitatively different deviation from the non-interacting behaviour: the Energy Shifts will stop us from taking the simplification used between (\ref{eq:IdealEvolution1RDMstates}) and (\ref{eq:IdealEvolution1RDMenergies}), while the Eigenstate Deformations forbid the step taken from (\ref{eq:IdealEvolution1RDMnone}) to (\ref{eq:IdealEvolution1RDMstates}).\\

\begin{table}[]
\begin{tabular}{|l|c|c|}
\hline
 & $\Delta_{\vec n}=0$ & $\Delta_{\vec n}\neq 0$\\ \hline
 $|\phi_{\vec n} \rangle=0$ &  $\hat H_0$ & $\hat H_D$ \\ \hline
  $|\phi_{\vec n} \rangle\neq 0$  & $\hat H_S$ & $\hat H$ \\ \hline
\end{tabular}\caption{A schema of the strategy followed to explore the role of Energy Shifts $\Delta_{\vec n}$ and Eigenstate Deformations $|\phi_{\vec n}\rangle$. We build two different Hamiltonians $\hat H_S$ and $\hat H_D$ combining properties from the interacting Hamiltonian $\hat H = \hat H_0 + \hat V$ and its non-interacting counterpart $\hat H_0$.}\label{tab:EDandES}
\end{table}

According to everything we have presented so far, it makes sense to ask what are the roles that the Energy Shifts and the Eigenstate Hamiltonians play individually. To answer this we will follow the strategy described in Table \ref{tab:EDandES}. We design artificial N-body Hamiltonians that show each one of these deviations from the ideal case in an isolated way. They will be the {\it Shifted Hamiltonian} $\hat H_S$ and the {\it Deformed Hamiltonian} $\hat H_D$\\

\begin{eqnarray}\label{def:HDandHS}
\hat H_S &=& \sum_{\vec k\, /\, N} \hbar \omega_{\vec k} \, |\vec k\rangle \langle \vec k |\\
\hat H_D &=& \sum_{\vec k\, /\, N} \hbar \vec k \cdot \vec \omega^0 \, |\omega_{\vec k}\rangle \langle \omega_{\vec k} |
\end{eqnarray}

\begin{center}
\begin{figure}
\includegraphics[width=0.5\textwidth]{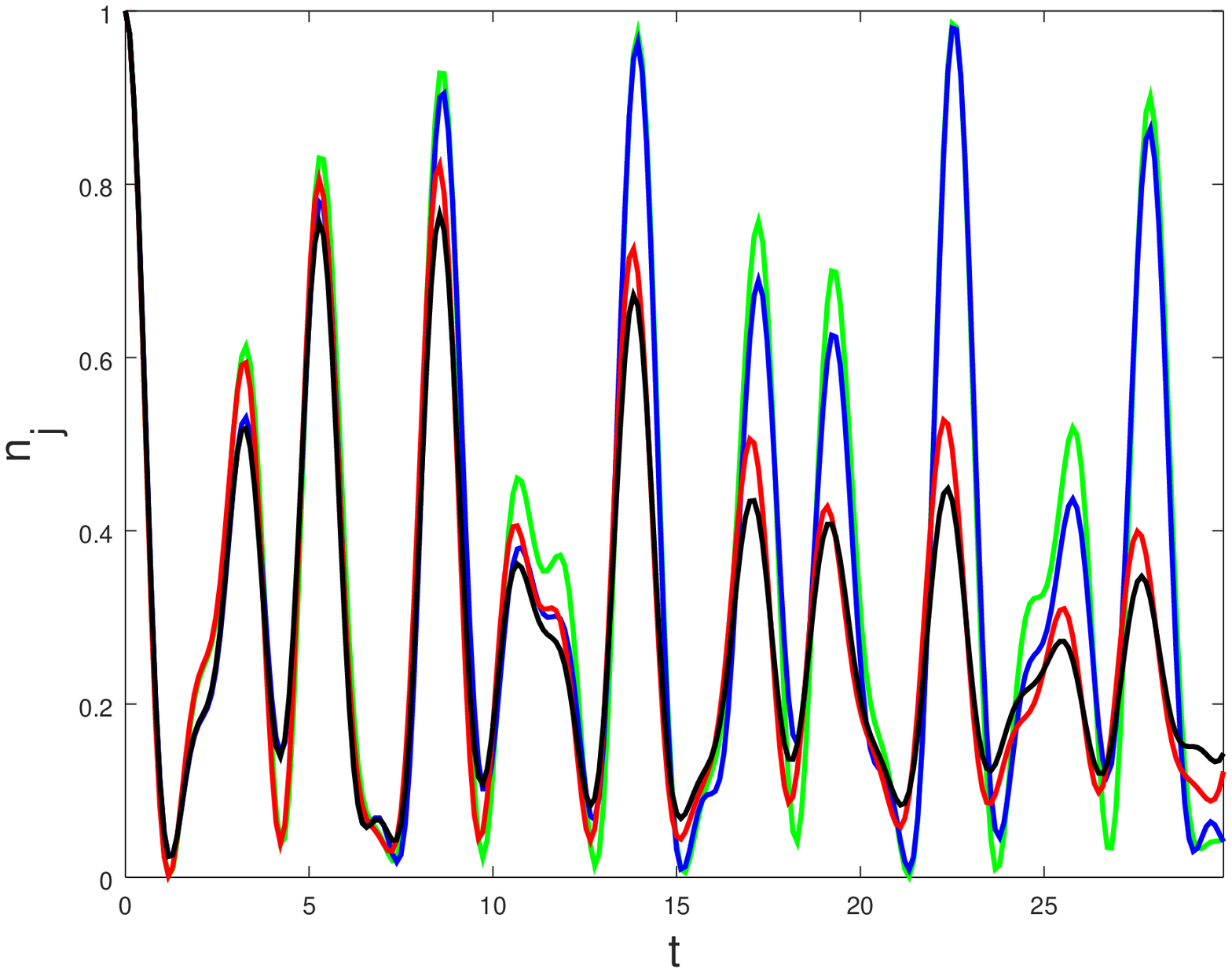}\caption{Compared evolution of the occupation of the site initially populated.$n_j(t=0)=1$ when we evolve under the Energy Shift (red) and using the Eigenstate Deformation (blue) Hamiltonians. At early stages of the relaxation process both approaches seem similar}\label{fig:CompareEXESEDoccupationRelax}
\includegraphics[width=0.5\textwidth]{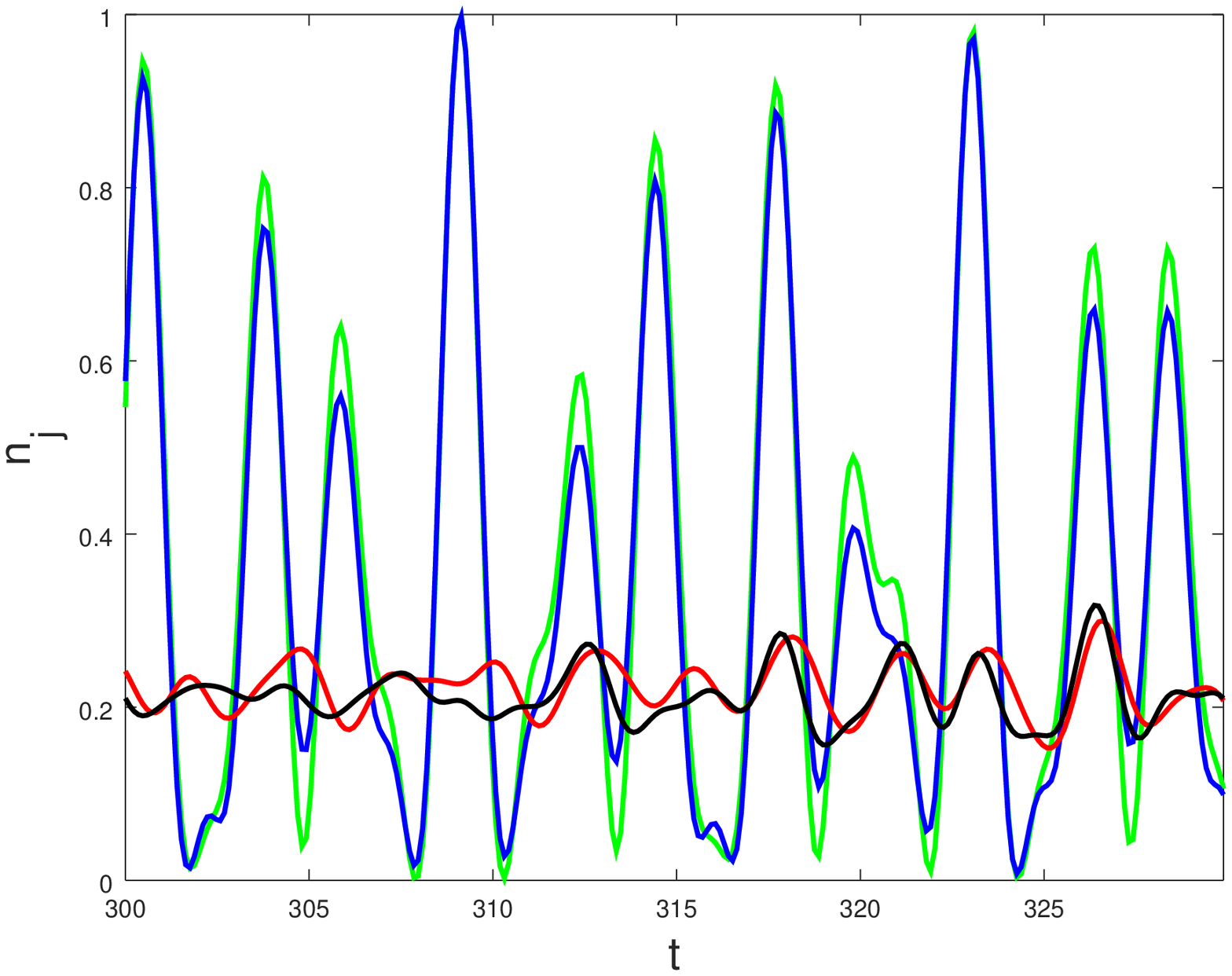}
\caption{Same as Fig.~\ref{fig:CompareEXESEDoccupationRelax} but after a reasonably long time. After a long enough evolution time only the Energy Shift Hamiltonian yields qualitatively realistic results.}\label{fig:CompareEXESEDoccupationSteady}
\end{figure}
\end{center}

The compared effect of time evolution under $\hat H_D$ or $\hat H_S$ can be seen in Fig.~\ref{fig:CompareEXESEDoccupationRelax} y Fig.~\ref{fig:CompareEXESEDoccupationSteady}, where we display the time evolution of the occupation number of the initially populated lattice site according to the four different Hamiltonians. In Fig.~\ref{fig:CompareEXESEDoccupationRelax} we see that for short times both $\hat H_D$ and $\hat H_S$ yield similar qualitatively similar results. This  is because the both capture the essential component of the evolution, which is nothing but `free flight', i.e. the dynamics provided by the non-interacting Hamiltonian. Still, once a reasonably long time has passed (see Fig.~\ref{fig:CompareEXESEDoccupationSteady}), the real system behaves in a way that we could call closer to ``equilibrium'' in the sense explained in previously in Section \ref{subsection:TimeAveraging}: a low and stable average value and small fluctuations around it. This long term dynamics is reasonably well described by $\hat H_S$ but not so by $\hat H_D$, that looks more similar to the evolution given by the ideal Hamiltonian.\\
 
Nevertheless, evaluating the time-evolved matrix elements we can assert that the Shifted Hamiltonian has one important weakness

\begin{eqnarray}\label{eq:Evolution1RDMwithHS}
\rho_1^{S}(p|q) &=& e^{-i t (\omega_p^0 - \omega_q^0)}\, \sum_{\vec n \, /\, N-1,M} e^{-it(\Delta_{\vec n p}-\Delta_{\vec n q})}\nonumber\\
& & \frac{\sqrt{(n_p+1)(n_q+1)}}{N} \langle \vec n p | \Psi_N^0\rangle \langle \Psi_N^0 | \vec n q \rangle\nonumber\\
& & 
\end{eqnarray}

The occupation numbers in the one-body energy representation do not evolve under the Shifted Hamiltonian because $[\hat H_S,\hat H_0]=0$. This does not happen for the Deformed Hamiltonian, where

\begin{eqnarray}\label{eq:MomentumOccupationNumbersED}
\rho_1^{D}(p|q) &=& \sum_{\vec m, \vec m^\prime / N} e^{-it(\vec m-\vec m^\prime)\cdot \vec \omega^0}\, R(\vec m, \vec m^\prime; p,q)\\
R(\vec m, \vec m^\prime; p,q) &=& \sum_{\vec n \, /\, N-1,M} \frac{\sqrt{(n_p+1)(n_q+1)}}{N} \nonumber\\
& & \times\, \langle \omega_{\vec m}|\Psi_N^0\rangle \langle \Psi_N^0| \omega_{\vec m^\prime}\rangle \,\langle \vec n p |\omega_{\vec m}\rangle \, \langle \omega_{\vec m^\prime}| \vec n q \rangle\nonumber\\
& & 
\end{eqnarray}

because in this case  $\langle  \omega_{\vec m}| \vec n p \rangle \neq \delta_{\vec m,\vec n p}$.\\
If our goal was to obtain the values of physical observables from the reduced representation once the stationary state has been achieved, then the description offered by $\hat H_D$ would be more interesting because it keeps more detailed information about the components of the initial state that will remain stationary. In turn, if we are interested about the description of the mid- and long-term dynamics then $\hat H_S$ yields qualitatively more realistic results. We can glimpse the reason when we compare what makes (\ref{eq:Evolution1RDMwithHS}) different from (\ref{eq:MomentumOccupationNumbersED}).\\
In the first one (\ref{eq:Evolution1RDMwithHS}) the deviations from the ideal dynamics given by $\hat H_0$ come from phase factors that depend only on the differences between Energy Shifts $\Delta_{\vec n p}-\Delta_{\vec n q}$, and since these are relatively small in comparison to other phase changes they will produce changes that are much slower than the non-interacting evolution. But because such phase factors do affect each and every one of the components $\langle \vec n p | \Psi_N^0\rangle$, this implies that in the long run the differences with respect to ideal evolution will not be limited in amplitude.\\
On the contrary, the differences from ideal evolution and the evolution described by the Deformed Hamiltonian (\ref{eq:MomentumOccupationNumbersED}) will have different components because now $\langle  \omega_{\vec m}| \vec n p \rangle = \delta_{\vec m,\vec n p} + \phi_{\vec m,\vec n p}$, with $\phi_{\vec m,\vec n p}\neq 0$. The component with the highest amplitude in (\ref{eq:MomentumOccupationNumbersED}) will follow from the terms where $\delta_{\vec m,\vec n p} \delta_{\vec m^\prime,\vec n q}$. But this component is precisely the equivalent to non-interacting evolution. Any deviation from ideal evolution will be given by the rest of the terms, all of them proportional in amplitude $\phi_{\vec m,\vec n p}$ o $\phi_{\vec m^\prime,\vec n q}$. This means that the deviations from ideality provided by $\hat H_D$ will be limited by the amplitude of the Eigenstate Deformation, that we could quantify through

\begin{eqnarray}\label{eq:TotalEigenstateDeformation}
D &=&\mbox{Max} \left[|\phi_{\vec n,\vec m} |^2\right]_{\vec n, \vec m \,/\, N,M}
\end{eqnarray}

\begin{table}[]
\begin{tabular}{|c|c|c|c|}
\hline
 N \,/\, M & 4 & 5 & 6 \\ \hline
 4 & >1.0e-45 & 9.0e-06 & 3.0e-05 \\ \hline
 5 & 8.9e-07 & 7.0e-05 & 5.6e-05 \\ \hline
 6 & 1.3e-05 & 1.5e-04 & 1.1e-04\\ \hline
\end{tabular}\caption{Estimation of the Eigenstate Deformation $D$ for the system sizes simulated, for $V=1/4$. We see that they remain at low values in all cases.}\label{tab:EigenstateDeformation}
\end{table}

In our numerical simulations this parameter is not larger than $O(10^{-4})$, as we can see in Table \ref{tab:EigenstateDeformation}. We can consider that this quantity will be always small as long as we an consider that the interactions are weak. Furthermore, considering cases beyond this scenario would not be consistent with our convention of using the same labelling $|\omega_{\vec n}\rangle$ for the eigenstates of both the ideal and the interacting Hamiltonian because our initial claim, i.e. that we can do this because the deformed eigenstates are very close to the non-deformed ones, will not be true any longer.\\

We are going to use a similar argumentation to analyse the evolution of quantum purity of the reduced representation. But since the non-interacting Hamiltonian conserves purity perfectly the deviations from ideal behaviour will be even more apparent.

\subsection{Evolution of quantum purity}\label{subsection:PurityEvolution}

\begin{center}
\begin{figure}
\includegraphics[width=0.5\textwidth]{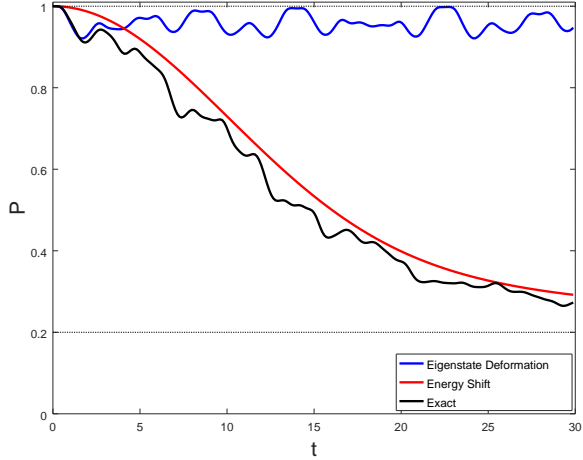}
\caption{Compared evolution of the purity of the one-body reduced density matrix when we evolve under the Energy Shift (red) and using the Eigenstate Deformation (blue) approximations. Even for systems of relatively small size ($N=M=5$ is represented, interaction strength $V=1/4$) the purity decays to values close to the minimum. Only the Energy Shift Hamiltonian predicts this decay.}\label{fig:PurityRelaxationEXES}
\end{figure}
\end{center}

While the bahaviour of all the four Hamiltonians is similar in the previous context, i.e. refered to the dynamics of quantum observables in the reduced representation (in the formal sense of observable), this is in part because the core of the dynamics was still what we informally called `free flight', this is, dynamics in the absence of interactions as given by $\hat H_0$. This changes drastically once we focus on the evolution of quantum purity, which is not an observable in the formal sense, and remains constant under such `free flight'.\\
We can see that this is true in Fig.~\ref{fig:PurityRelaxationEXES}, where the evolution of the quantum purity under the different Hamiltonians is represented. The system departs from a state that is pure $P(t=0)=1$ in its coarse-grained representation. The evolution under $\hat H_0$ keeps purity constant but the differences between $\hat H_D$ (blue) and $\hat H_S$ (red) are now much more evident, being still $\hat H_S$ the one that describes better the dynamics of the actual Hamiltonian $\hat H$ (black).

The differences between them have an origin similar to that discussed in the previous section. $\hat H_D$ fluctuates fast and with limited amplitude, while $\hat HS$ still has phase factors with much smaller frequencies - therefore slower - but its fluctuations are unbounded in amplitude. Since quantum purity is - among all clues of coherence - the simplest to calculate analytically, we can use the previous result (\ref{eq:Evolution1RDMwithHS}) to explore this behaviour. Then we can write for the Shifted Hamiltonian

\begin{eqnarray}\label{eq:PurityEvolutionHS}
P(t)&=&\sum_{p,q} | \rho_1(p|q) |^2 \nonumber \\
&=& \sum_{p, q} \sum_{\vec n, \vec m\,/\, N-1,M} \, e^{-it(\Delta_{\vec n p}-\Delta_{\vec n q}-\Delta_{\vec m p}+\Delta_{\vec m q})}
\nonumber\\
& & \times \,\frac{\sqrt{(n_p+1)(n_q+1)(m_p+1)(m_q+1)}}{N^2} \nonumber\\
& & \times \,\langle \vec n p|\Psi_N^0\rangle \langle \Psi_N^0 | \vec n q \rangle\,\langle \Psi_N^0 | \vec m p\rangle \langle \vec m q|\Psi_N^0 \rangle\nonumber\\
& & 
\end{eqnarray}

Consistently with the statement made, that our conception of Self-Induced Decoherence is based on the same principle of time-averaging and limited fluctuations borrowed from Statistical Mechanics, it makes sense to split the terms from the expression above in its stationary and fluctuating components

\begin{eqnarray}
P(t) &=& P_{AV} + \delta P(t)
\end{eqnarray}

where the stationary value $P_{AV}$ is the sum of all terms where the phase factors cancel (either because $p=q$ or else because $\vec n =\vec m$)

\begin{eqnarray}\label{eq:PurityBaseValueHS}
P_{AV} &=& \sum_{p,q}\sum_{\vec n \, /\, N-1}\frac{(n_p+1)(n_q+1)}{N^2}\,|\langle \vec n p | \Psi_N^0 \rangle |^2 |\langle \Psi_N^0 | \vec n q \rangle|^2\nonumber \\
& & + \sum_p \sum_{\vec n \neq \vec m} \frac{(n_p+1)(m_p+1)}{N^2}\left| \langle \vec n p|\Psi_N^0\rangle \right| ^2\left| \langle \vec m p|\Psi_N^0 \rangle \right| ^2\nonumber \\
& & 
\end{eqnarray}

The fluctuations $\delta P(t)$ correspond to the terms where phase factors do not cancel ($p\neq q$ and $\vec n \neq \vec m$)

\begin{eqnarray}\label{eq:PurityFluctuationsHS}
\delta P(t) &=& \sum_{p \neq q} \sum_{\vec n \neq \vec m\,/\, N-1,M} \, e^{-it(\Delta_{\vec n p}-\Delta_{\vec n q}-\Delta_{\vec m p}+\Delta_{\vec m q})}\, \nonumber\\
& &\times\, \frac{\sqrt{(n_p+1)(n_q+1)(m_p+1)(m_q+1)}}{N^2}\nonumber\\
& & \times \, \langle \vec n p|\Psi_N^0\rangle \langle \Psi_N^0 | \vec n q \rangle\,\langle \Psi_N^0 | \vec m p\rangle \langle \vec m q|\Psi_N^0 \rangle \nonumber\\
& &
\end{eqnarray}

\begin{center}
\begin{figure}
\includegraphics[width=0.5\textwidth]{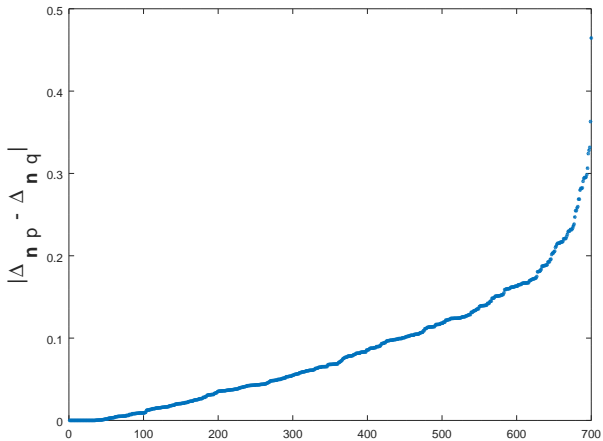}
\caption{Differences between the energy shifts $|\Delta_{\vec n p}-\Delta_{\vec n q}|$ involved in each one of the phase factors in the sum (\ref{eq:Evolution1RDMwithHS}) that yields the time evolution of the reduced density matrix elements $\rho^S(p|q)$ under the Energy Shift Hamiltonian $\hat H_S$. The size of the Hilbert space for $N=M=5$ is $L_{5,5}= 126$, meaning that up to $7.875$ pairs of shift differences $(\vec m,\vec m^\prime)$ could be involved. But only the shift differences of the form $(\vec m,\vec m^\prime)=(\vec n p,\vec n q)$, $p \neq q$ are relevant (only $700$ possibilities).}\label{fig:DifferenceOfEnergyShifts}
\end{figure}
\end{center}

Our main interest concerning the fluctuations it to determine if their amplitude is going to remain bounded most of the time. If the contributing terms add constructively very often we would sporadically observe sudden peaks of high quantum purity, something that would not fit our concept of decoherence.\\
Our plan implies to consider that terms with phase factors of different frequencies can be treated as independent random variables. This will require those frequencies to respect certain conditions: they should of course be different from each other; they should not be an integer multiple of each other (i.e. they are incommesurable with respect to each other); and the differences between them should not be so small that we would not be able to distinguish them on a reasonable observational time scale (see the Appendix \ref{appendix:SumOscillating}). In practice this hypothesis may not always be fulfilled altogether. But this is not the case of the concrete model that we are using as an example, as we can see represented in Fig.~\ref{fig:DifferenceOfEnergyShifts}. From all terms with phase factors dependent on the differences between Energy Shifts $|\Delta_{\vec m} - \Delta_{\vec m^\prime}|$ that could contribute to the fluctuations $\delta P(t)$, all those that are not of the form $(\alpha,\beta)=(\vec n p, \vec n q)$ will have phase factors that cancel and these terms contribute to the average value instead, just as we expected - but among those that do have this form (represented in the figure) only a few are small enough to be considered non-fluctuating terms. If we take the simulation time represented in Fig.~\ref{fig:PurityRelaxationEXES} as our reference time scale then the threshold to discard the lowest values is $1/T\approx 0.033$, leaving 503 (72$\%$) of them above it. \\


\begin{center}
\begin{figure}
\includegraphics[width=0.5\textwidth]{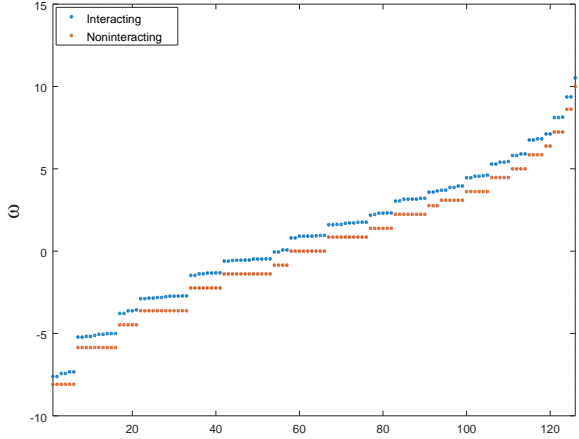}
\caption{Energy spectrum of the Bose-Hubbard Hamiltonian for $N=M=5$ and $V=1/4$. We see that the interaction displaces the energy levels (blue) from the non-interacting ones (red). These differences are the Energy Shifts $\Delta$}\label{fig:EnergyShifts}
\end{figure}
\end{center}

\begin{center}
\begin{figure}
\includegraphics[width=0.5\textwidth]{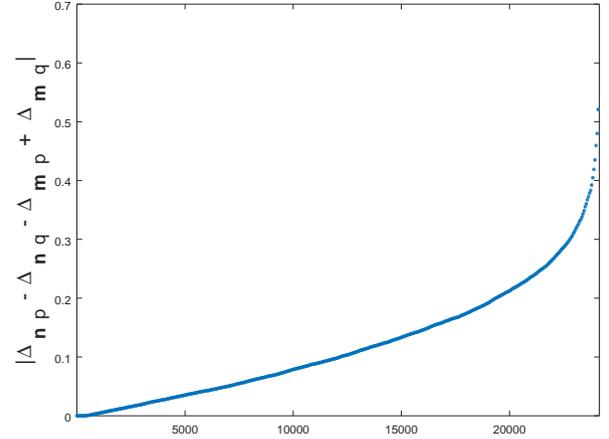}
\caption{Differences between the difference between energy shifts $|\Delta_{\vec n p}-\Delta_{\vec n q} - \Delta_{\vec m p} + \Delta_{\vec m q}|$ involved in each one of the phase factors in the terms contributing to purity fluctuations (\ref{eq:PurityFluctuationsHS}). The amount of shift differences previously calculated amounted to $700$, meaning that up to $244.650$ terms with different frequencies could be involved. But again, only the differences of the form $(\vec n p,\vec n q,\vec m p,\vec m q)$, $\vec n\neq\vec m$, $p \neq q$ are relevant (only $24.150$ possibilities). We have represented in the picture only a sample of them (one out of $50$).}\label{fig:Diff_Of_Diff}
\end{figure}
\end{center}

Notice that the differences represented in Fig.~\ref{fig:DifferenceOfEnergyShifts} $|\Delta_{\vec n p}-\Delta_{\vec n q}|$ will be crucial when distinguishing stationary and fluctuating terms - in the evolution of $\rho^S(p|q)$, just as we expressed in (\ref{eq:Evolution1RDMwithHS}). For the evolution of quantum purity it becomes more complicated because the phase factors depend instead on the differences between the differences between the Energy Shifts $|\Delta_{\vec n p}-\Delta_{\vec n q} - \Delta_{\vec m p} + \Delta_{\vec m q}|$. They are represented in Fig.\ref{fig:Diff_Of_Diff}, obtaining similar results. The principles explained so far are still the same and the main hypothesis is that the resulting frequencies should not cancel or be redundant too often.  If we apply the same criterion explained above (time scale from Fig.~\ref{fig:PurityRelaxationEXES} as reference) then 19.389 (80$\%$) of the terms would contribute to the fluctuations.\\

As long as this `incommesurability conjecture' is acceptable we can suppose that each oscillating term is a random variable independent from the others and we can calculate the variance of the sum of all of them as

\begin{eqnarray}\label{eq:Sigma2PurityHS}
\bar \sigma^2_P&=& \sum_{p \neq q} \sum_{\vec n \neq \vec m} \frac{(n_p+1)(n_q+1)(m_p+1)(m_q+1)}{N^4} \nonumber\\
& & \times \,\left| \langle \vec n p|\Psi_N^0\rangle \langle \Psi_N^0 | \vec n q \rangle\,\langle \Psi_N^0 | \vec m p\rangle \langle \vec m q|\Psi_N^0 \rangle \right| ^2\nonumber \\
& & 
\end{eqnarray}

The two results above (\ref{eq:PurityBaseValueHS}) and (\ref{eq:Sigma2PurityHS}) can be easily checked in our test system numerically - a Bose-Hubbard Hamiltonian departing from an initial state (\ref{eq:InitialStateSingleModeBH}) with all particles localized in one lattice site at position $x$. This state can be rewritten as

\begin{eqnarray}\label{eq:InitialStateSingleMode}
|\Psi_N^0\rangle &=& \sum_{\vec n / N, M} \sqrt{\frac{N!}{n_1!...n_M!}}\langle 1|x \rangle^{n_1}...\langle M|x \rangle^{n_M}\, |\vec n\rangle\nonumber\\
& & 
\end{eqnarray}

Notice that $x$ is an eigenstate of `position' (lattice site) representation, while the indexes in the sum $\vec n = (n_1,n_2,...,n_k,...,n_M)$, with $k=1,2,...,M$ are in the `quasimomentum' (one-body energy) eigenstates.\\
This initial state in particular is very simple because the components of a localized state have all the same amplitude

\begin{eqnarray}\label{eq:ScalarProductPositionMomentumBH}
|\langle x | k \rangle|^2 &=& 1/M
\end{eqnarray}

and this simplifies enormously the calculation. For the average value (\ref{eq:PurityBaseValueHS}) we obtain

\begin{eqnarray}\label{eq:PurityBaseValueSSPISAnsatz}
P_{AV} &=& \frac{1}{M}+M(M-1)\frac{a_{N-1}^{(M)}}{M^{2N}}
\end{eqnarray}

and for the variance of fluctuations(\ref{eq:Sigma2PurityHS}) it is

\begin{eqnarray}\label{eq:Sigma2PuritySSPISAnsatz}
\bar \sigma^2_P &=& \frac{M(M-1)}{2}\left[ \left(\frac{a_{N-1}^{(M)}}{M^{2N}} \right)^2 - \frac{b_{N-1}^{(M)}}{M^{4N}} \right]
\end{eqnarray}

where both results are expressed in terms of sums of powers of the multinomial series

\begin{eqnarray}\label{eq:MultinomialPowerSum}
a_n^{(m)} &=& \sum_{\vec k / n, m} {{n}\choose{\vec k} }^2\\
b_n^{(m)} &=& \sum_{\vec k / n, m} {{n}\choose{\vec k} }^4\\
{{n}\choose{\vec k} }&=&\left(\frac{n!}{k_1!k_2!...k_m!} \right)
\end{eqnarray}

\begin{center}
\begin{figure}
\includegraphics[width=0.5\textwidth]{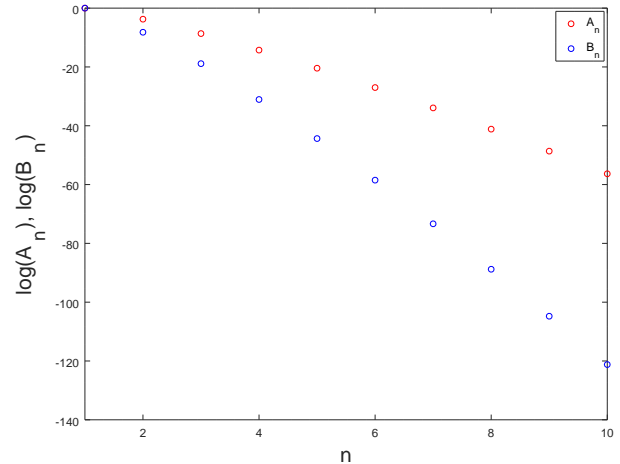}
\caption{Behaviour of the coefficients $A_n=n^{-2n}\,a_n^{(n)}$ (red) and $B_n=n^{-4n}\,b_n^{(n)}$ (blue), on which the variance of purity fluctuations $\bar \sigma_P^2$ depend when we evolve under the Shifted Hamiltonian $\hat H_S$.}\label{fig:AnBnCoefficients}
\end{figure}
\end{center}

This result allows us to have an idea of how fast the amplitude of purity fluctuations decay as we consider systems of larger sizes. For example, in Fig.~\ref{fig:AnBnCoefficients} we can see that the terms on which the variance depends $A_n=n^{-2n}\,a_n^{(n)}$ y $B_n=n^{-2n}\,b_n^{(n)}$ decay at least as  $\log{A_n} \propto n-1$ (same for $B_n$). The amplitude of purity fluctuations would be notoriously limited, as we expected.

\section{Conclusions}
\label{Conclusions}

In this work we have recalled and supported with numerical evidence the concept of Self-Induced Decoherence for the coarse-grained description of quantum systems. Although this concept is not new it has been mostly underestimated in the previous literature about quantum decoherence. While it is very hard to consider real systems as perfectly isolated, the effect is present and should not be disregarded.\\

The concept of Self-Induced Decoherence exposed here requires us to remember the basic principles of Pure-State Quantum Statistical Mechanics. This is, that our access to the system is limited to a simplified representation of it, based on a few-body (usually one body) Reduced Density Matrix. That when we refer to a `stationary state' of a many-body system after undergoing a relaxation process, it does not mean that the system actually remains stationary, but that the accessible quantities remain {\it at almost any time} around well defined and stable average values. And that the deviations from such average values are only fluctuations of an amplitude that, on average, is extremely small and proportionally less relevant as we consider systems of larger sizes.\\

Keeping these concepts in mind and accepting for the sake of simplicity that coherent quantum systems will present a higher quantum purity than incoherent ones we could verify indeed that the coarse-grained  representation of a many-body system of isolated interacting particles loses quantum coherence. We added a qualitative analysis of the causes that make interacting Hamiltonians to lose quantum coherence in the way previously described. It is based on splitting the two obvious characteristics that distinguish an interacting Hamiltonian from its non-interacting counterpart: the Eigenstate Deformation and the Energy Shifts. Our analysis suggests that the first one (small in amplitude but fast) will play a major role in fluctuations, while the second one (large in amplitude but slow) is the responsible for the long time stationary behaviour of quantum purity.\\

About the aforementioned fluctuations, we could make an analytical estimation of its amplitude based on the Energy Shift Approximation and taking an initial state - the fully localized state - that is uniformly distributed among the non-interacting energy spectrum. We could see that this amplitude decreases extremely fast as the size of the system grows. But the actual amplitude of fluctuations will not be that small for two clear reasons. The first one, more obvious, is that the Eigenstate Deformations will add a significant contribution to the purity fluctuations. And the second one, more subtle, is that some of the oscillation frequencies involved in the fluctuating terms of the quantum purity may be either redundant - and therefore result in stationary contributions - or else so similar that they will not be distinguishable in a reasonable observational time scale. The fact that redundancies - in fact, the lack of them - in the N-body energy spectrum plays an important role in relaxation of isolated quantum systems has been already pointed out both explicitly \cite{Reimann2012,Short2012} and implicitly in relation to the Eigenstate Thermalization Hypothesis \cite{Rigol2012,Rigol2016}. But we could observe that for quantum purity this principle becomes slightly more complex because it appeals to the (non-)redundancy of differences of differences of energy shifts. \\

We have put under test all of these concepts in a system simple enough to admit exact solutions - the few-body Bose-Hubard model. With this we have provided numerical evidence supporting that the principles exposed are fulfilled in this case, even for very limited system sizes. For this system and this particular initial conditions we cannot talk properly about relaxation to a stationary state because we can see that observables (local occupation numbers, in particular) do not stop to oscillate even after a long time. But even then we can see that the quantum purity, in turn, does decay to a very stable lower value, a fact that we could interpret as a tendency to incoherent behaviour of the coarse-grained accessible representation.\\

Concerning possible experimental implementations, we can consider ourselves lucky because the Bose-Hubbard model has had its laboratory counterpart for a very long time already, in the form of trapped bosons in a one-dimensional magneto-optical lattice. This is another reason why Self-Induced Decoherence in the Bose-Hubbard model has obvious practical implications. As an example, the interference patterns of the bosons released from a lattice would lose contrast depending on how long they have been stored before being released. Their lifetime as quantum coherent systems would be limited regardless of how well isolated from outer sources of decoherence we keep the sample \cite{Ott2004}.


\begin{appendix}

\section{Sum of oscillating quantities with different frequencies}\label{appendix:SumOscillating}

Consider an oscillatory variable $y$, oscillating with frequency $\Omega$

\begin{eqnarray}
y(t) &=& \sin{(\Omega t + \alpha)}
\end{eqnarray}

This non-random variable can be considered random is the instant $t$ when we measure its value is taken randomly from a sampling interval $T$. If this observation interval is exactly one single oscillation period then the measured values are distributed according to

\begin{eqnarray}
p(y) &=& \frac{1}{\pi}\frac{1}{\sqrt{1-y^2}}
\end{eqnarray}

having for average value and variance

\begin{eqnarray}
\bar y &=& 0\\
\bar \sigma_y^2 &=& \frac{1}{2}
\end{eqnarray}

If instead of a single oscillation period our samplig time $T$ covers an exact multiple of full oscillation periods the results will distribute in the exact same way. But even if $T$ does not cover an exact integer multiple of full periods the results will not differ much in comparison with those obtained for an exact integer multiple as long as $T$ covers many full oscillation periods.\\

If our measured variable $y$ were the sum of many oscillating terms

\begin{eqnarray}\label{eq:SumOfOscillatingVariables}
y(t) &=& \sum_{j} y_j\, \sin{(\Omega_j t + \alpha_j)}
\end{eqnarray}

the terms contributing to $y$ could not be treated as independent random variables if, for example, their frequencies $\Omega_j$ were equal or too close to be distinguished within the sampling time considered $|\Omega_i-\Omega_j|\ll \frac{2\pi}{T}$. \\

But as long as we can consider all the frequencies invoveld as different within our sampling time interval, then we could treat each term as an independent variable from the rest and we can calculate the average value and variance of (\ref{eq:SumOfOscillatingVariables}) as

\begin{eqnarray}
\bar y &=& 0\\
\bar \sigma_y &=& \frac{1}{2}\sum_{j}y_j^2
\end{eqnarray}

\end{appendix}


\bibliographystyle{apsrev}

\end{document}